\documentclass[%
aps,
jmp,%
amsmath,amssymb,
reprint,%
]{revtex4-2}
\usepackage{graphicx}
\usepackage{subfigure}
\usepackage{dcolumn}
\usepackage{bm}
\usepackage[mathlines]{lineno}
\usepackage{graphicx}

\usepackage{graphicx}
\usepackage{mathptmx}
\usepackage{subfigure}
\usepackage{dcolumn}
\usepackage{bm}
\usepackage[mathlines]{lineno}
\graphicspath{{figure/}}
\usepackage{array}
\usepackage{braket}
\usepackage{diagbox}
\usepackage{amsmath}
\usepackage{appendix}
\usepackage{multirow}
\usepackage{bm,color,bbm}

\newcommand{\blk}{\color{black}}

\usepackage{float}
\newcolumntype{.}{D{.}{.}{-1}}
\usepackage{amsopn}
\usepackage[colorlinks,linkcolor=blue, anchorcolor=blue, urlcolor=blue, citecolor=blue]{hyperref}
\usepackage{epstopdf}

\begin{document}

	\title {Refrence-Frame-Independent Quantum Key Distribution over \\ 250 km of Optical Fiber}
	\author{Xin Liu$^{1}$, Di Luo$^{1}$, Zhicheng Luo$^{1}$, Shizhuo Li$^{1}$, Zhenrong Zhang$^{2}$, and Kejin Wei$^{1,*}$}
	\address{
		$^1$Guangxi Key Laboratory for Relativistic Astrophysics, School of Physical Science and Technology,
		Guangxi University, Nanning 530004, China\\		$^2$Guangxi Key Laboratory of Multimedia Communications and Network Technology, School of Computer Electronics and Information, Guangxi University, Nanning 530004, China\\
		$^*$Corresponding author: kjwei@gxu.edu.cn
	}

	\begin{abstract}
	The reference-frame-independent quantum key distribution (RFI QKD) protocol enables QKD systems to function effectively despite slowly varying reference frames, offering a distinct advantage in practical scenarios, particularly in mobile platforms. In this study, we successfully   distribute secure key bits over a 250 km optical fiber distance by developing an RFI QKD system with a repetition rate of 150 MHz. Benefiting  from high repetition rate, we achieve a  finite-key secret key rate of 49.65 bit/s at a distance of 200 km, which is more than three times higher than state-of-the-art systems. Our work dramatically extends the transmission distance and enhances the  secret key rate of RFI QKD, significantly promoting its practical application.
	\end{abstract}
	\maketitle
	
	\section{Introduction}
	
	Quantum key distribution (QKD) enables the sharing of information-theoretically secure keys between two remote legitimate communication parties, commonly referred to as Alice and Bob, even in the presence of malicious eavesdroppers.
	Since the proposal of the first QKD protocol in 1984~\cite{BB84}, significant advancements have been made in QKD concerning   transmission distance~\cite{Lo2012,Lucamarini2018,Boaron2018,Liu2023,Zhou2023}, secret key rates (SKRs)~\cite{Wang2019asymmetric,Jiang2022,Grunenfelder2023,Li2023}, miniaturization~\cite{Paraiso2019,Zhang2019integrated,Beutel2022,Wei2023}, and network structures~\cite{Wengerowsky2018,Chen2021,Fan-Yuan2022,Avesani2022}.
	
	In most QKD systems, the calibration of reference frames is a crucial step that affects the final quantum bit error rate (QBER). However, instabilities in the transmitter and receiver, as well as unstable communication channels, can cause time-varying misalignment of reference frames,  resulting in a higher QBER. Specifically, for the phase-encoded protocol, the time-varying phase difference between Alice and Bob, affects the interferometric stability. To maintain system operation, Alice and Bob typically need additional devices to compensate for the drift of reference frames in   real time~\cite{Tang2023,Pathak2023}, thereby increasing system complexity.
	
	The reference-frame-independent QKD (RFI QKD) protocol, first proposed by Laing et al.~\cite{Laing2010}, enables Alice and Bob to share secure keys despite slowly varying reference frames, thereby eliminating the need for phase feedback devices and significantly reducing system complexity. This unique property of RFI QKD has garnered significant attention from researchers. Over the past decades, significant progress has been made in extending transmission distance and key rates~\cite{Xue2020,Zhu2022,Jiang2023}, relaxing technical challenges~\cite{Zhang2014,Tannous2019,Li2020}, and closing realistic security loopholes~\cite{Sun2021,Lim2021,Nie2022}. Notably, RFI measurement-device-independent QKD (RFI MDI QKD), which can eliminate all side-channel loopholes in measurement devices, has been proposed~\cite{Yin2014} and its practicality has been demonstrated~\cite{Wang2015,Lee2020,Lu2020,Jin2021,Zhou2021,Liu2021,Wei2022,Zhu2023}.	
	
	Our primary focus is on standard decoy-state RFI QKD, which has been extensively studied theoretically and experimentally in   fiber~\cite{Liu2019,Tang2022}, free-space~\cite{Li2019,Yoon2019,Xue2020,Chen2020}, and underwater~\cite{Sun2021Performance} channels, with the potential for initial large-scale deployments. However, due to the finite-size effect caused by the limited number of sent light pulses and the high dark-count rate of single-photon detectors, the record-breaking transmission distance of RFI QKD is 175 km at a repetition rate of 100 MHz~\cite{Tian2024}. This is significantly lower than the more mature BB84 QKD, which operates at a GHz repetition rate and achieves a longer distance of 421 km~\cite{Boaron2018}. \blk 
	
	In this study, we successfully implement an RFI QKD system with a repetition rate of 150 MHz. In the experimental test, we successfully distribute secure key bits over fiber channels up to 250 km with a misalignment of reference frames of approximately   $\theta=\pi /9 $. Due to the high repetition rate, an SKR of 49.65 bit/s is achieved over a 200-km optical fiber distance, which is 3 times higher than that of the state-of-the-art system reported in Ref.~\cite{Tian2024}. This work significantly extends the transmission distance and enhances the SKR of RFI QKD, further facilitating its practical application.
	
	The remainder of this paper is organized as follows. In Sec.~\ref{Protocol}, we briefly summarize the one-decoy state RFI QKD protocol adopted in the experiment. In Sec.~\ref{Experimental setup}, we  provide a detailed description of the setup for conducting RFI QKD experiment. Subsequently, in Sec.~\ref{Experimental results}, the experimental results are presented and discussed. Finally, we summarize the experimental results and discuss next steps in Sec.~\ref{Conclusions and discussion}. 
	
	\section{Protocol}
	\label{Protocol}
	 \begin{figure*}[!t]
		\centering
		\includegraphics[width=0.8 \linewidth]{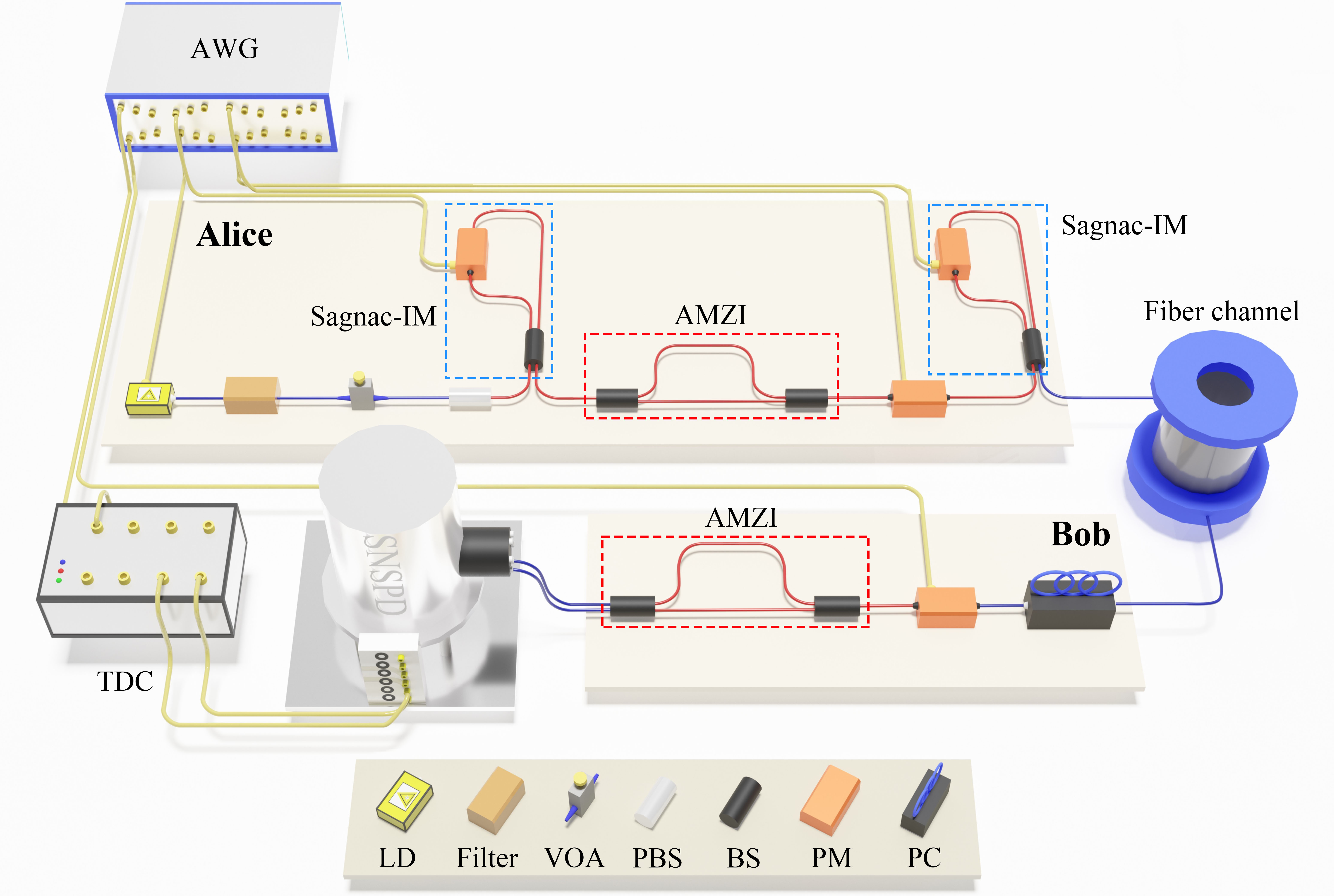}
		\caption{Experimental setup. LD: laser diode with a wavelength of 1550.05 nm; Filter: 10-GHz bandwidth filter; VOA: variable optical attenuator; PBS: polarization beam splitter; BS: polarization-maintaining beam splitter; PM: phase modulator; the blue dotted frame: Sagnac-based intensity modulator (Sagnac-IM); the red dotted frame: asymmetric Mach-Zehnder interferometer (AMZI); Fiber channel: standard commercial fiber spool; PC: polarization controller; SNSPD: superconducting nanowire single photon detector; AWG: arbitrary waveform generator; TDC: time-to-digital converter; the blue solid line: single mode fiber; the red solid line: polarization-maintaining fiber.}
		\label{RFI-QKD}
	\end{figure*}
	
	Firstly, we briefly summarize the one-decoy-state RFI-QKD protocol~\cite{Li2020} used in the experiment. In this protocol, Alice randomly prepares quantum states with intensity $k\in \left \{ \mu ,\nu  \right \} $ in three orthogonal bases $\alpha \in \left \{ Z_{A} , X_{A}, Y_{A} \right \} $ and transmits them to Bob via insecure quantum channels. Bob then randomly chooses a basis  $\beta \in \left \{ Z_{B}, X_{B}, Y_{B} \right \} $ to measure the received quantum states. The $\mathit{Z}$ basis is naturally aligned for Alice and Bob, while the misalignment exists in the $\mathit{X}$ and $\mathit{Y}$ bases (phase encoding) due to unavoidable time-varying phase drift. If the misalignment angle is denoted by $\theta$, the relationship between the reference frames of Alice and Bob can be expressed as
	\begin{equation}
		\begin{aligned}
			Z_{B} & =Z_{A}, \\
			X_{B} & =X_{A}\cos \theta + Y_{A}\sin \theta, \\
			Y_{B} & = Y_{A}\cos \theta - X_{A}\sin \theta.
		\end{aligned}
	\end{equation}     
	\blk
	Subsequently, using the well-aligned $\mathit{Z_{A}}$ and $\mathit{Z_{B}}$ bases, Alice and Bob reveal their chosen bases, distill the raw keys, and estimate the QBER, expressed as
	\begin{equation}
		\begin{aligned}
			E_{ZZ} =\frac{1-\left \langle Z_{A} Z_{B} \right \rangle }{2}.
		\end{aligned}
	\end{equation}
	Using the information collected from the $\mathit{X}$ and $\mathit{Y}$ bases, an intermediate quantity $C$ can be defined to estimate the amount of information leaked to Eve, which is specifically expressed as
	\begin{equation}
		\begin{aligned}
			C&=\left\langle X_{A} X_{B}\right\rangle^{2}+\left\langle X_{A} Y_{B}\right\rangle^{2}+\left\langle Y_{A} X_{B}\right\rangle^{2}+\left\langle Y_{A} Y_{B}\right\rangle^{2}  \\
			&= \left(1-2 e_{X X}^{1, U}\right)^{2}+\left(1-2 e_{X Y}^{1, U}\right)^{2} \\
			& +\left(1-2 e_{Y X}^{1, U}\right)^{2}+\left(1-2 e_{Y Y}^{1, U}\right)^{2}.
		\end{aligned}
	\end{equation}
	Note that the value of $C$ is independent of the reference frames misalignment angle $\theta$. The upper bound of information leakage due to Eve eavesdropping can be given by
	\begin{equation}
		\begin{aligned}
			I_{E}^{U} = \left ( 1-e_{Z Z}^{1,U} \right ) h\left ( \frac{1+u}{2} \right ) + e_{Z Z}^{1,U} h\left ( \frac{1+v}{2} \right ), 
		\end{aligned}
	\end{equation}
	where  $h\left ( x \right ) =-x\text{log} _{2} x-\left ( 1-x \right ) \text{log} _{2} \left (1- x \right )$ denotes the binary Shannon entropy, and 
	\begin{equation}
		\begin{array}{l}
			u=\text{min} \left [ \frac{\sqrt{C/2} }{1-e_{Z Z}^{1,U}},1 \right ],
		\end{array} 
	\end{equation}
	
	\begin{equation}
		\begin{array}{l}
			v= \frac{\sqrt{C/2-{{\left ( 1-e_{Z Z}^{1,U} \right )^{2}u^{2}}}}}{e_{Z Z}^{1,U}}. 
		\end{array} 
	\end{equation}
	
	Finally, Alice and Bob perform error correction and privacy amplification to distill the final secret keys. The secret key rate with finite-key analysis is calculated as
	\begin{equation}
		\begin{aligned}
			R & = \frac{1}{N_{t o t}}\left[s_{Z Z, 0}^{L}+s_{Z Z, 1}^{L}\left(1-I_{E}^{U}\right)-n_{Z Z} f h\left(E_{Z Z}\right)\right. \\
			&\left. - a \log _{2}\left(b / \varepsilon_{\mathrm{s e c}}\right)-\log _{2}\left(2 / \varepsilon_{\mathrm{c o r}}\right)\right],
		\end{aligned} \label{key rate}
	\end{equation}
	where $N_{tot}$ is the total pulse numbers sent by Alice. $s_{Z Z, 0}^{L}$ represents the lower bound of vacuum events, indicating Bob records a detection while the pulses sent by Alice contain no photons. $s_{Z Z, 1}^{L}$ represents the lower bound of single-photon events, defined as the number of detections by Bob when the pulses sent by Alice contain only one photon. $n_{Z Z}$ and $E_{Z Z}$ denote the length of the raw key and QBER when Alice sends $\mathit{Z}$ basis and Bob employs $\mathit{Z}$ basis measurement, respectively. $f=1.16$ is the error-correction efficiency. $a=6$ and $b=43$ depend on the specific security analysis taken into account. $\varepsilon_{\mathrm{sec}}=10^{-9}$ and $\varepsilon_{\mathrm{c o r}}=10^{-15}$ are the secrecy and correctness parameters, respectively. The derivation process of specific parameters can be found in Appendix~\ref{Appendix A}.
	
	\begin{table*}[]
		\centering
		\caption{ Comparison of recent RFI QKD experiments. ‘· · ·’ indicates that the misalignment angle $\theta$ is not considered.}
		\centering
		\fontsize{9}{12} \selectfont
		\begin{tabular}{ccccc}
			\hline \hline
			& Repetition rate (MHz) & Distance (km) & Misalignment angle $\theta$ & SKR (bit/s) \\ \hline
			Liang et al.~\cite{Liang2014} & 1        & 65         & · · ·       & 0.33     \\ \hline
			Liu et al.~\cite{Liu2019}     & 1        & 55         & $\pi /4$       & 8.36     \\ \hline
			Tang et al.~\cite{Tang2022}   & 80        & 150       & · · ·       & 10.31     \\ \hline
			Tian et al.~\cite{Tian2024}   & 100       & 175       & · · ·       & 16.40     \\ \hline
			\multirow{4}{*}{our work}     & \multirow{4}{*}{150}  &50 &\multirow{4}{*}{$\pi/9$}     & 189080.80   \\
			&           & 150       &                 & 1004.15   \\
			&           & 200       &                 & 49.65     \\
			&           & 250       &                 & 0.65     \\ 
			\hline \hline
		\end{tabular}
		\label{Comparison}
	\end{table*}
	
	\section{Experimental setup}
	\label{Experimental setup}
	
	The schematic of our setup, shown in Fig.~\ref{RFI-QKD}, comprises three main parts: transmitter (Alice), channel, and receiver (Bob).	
	Alice has a gain-switched laser diode (LD) that generates phase-randomized laser pulses with a   repetition rate of 150 MHz and a FWHM of about 140 ps. These generated light pulses are filtered through a 10-GHz bandwidth filter, resulting in a linewidth of approximately 32 pm. They are attenuated to a single-photon level using a variable optical attenuator (VOA). Since the filter and VOA are single-mode devices, a polarizing beam splitter (PBS) is utilized to purify the polarization of the light pulses. Next, the light pulses are coupled into a Sagnac-based intensity modulator (Sagnac-IM)~\cite{Roberts2018,Ma2021}, which includes a polarization-maintaining beam splitter (BS) and a phase modulator (PM).	The intensities of the light pulses are modulated by adjusting the applied voltages to generate the signal and decoy states.	
	
	The light pulses are subsequently divided into long ($l$) and short ($s$) signals upon entering an asymmetric Mach-Zehnder interferometer (AMZI) with a 2.5-ns delay. The following PM is used to prepare the states in the $\mathit{X}$ or $\mathit{Y}$ basis, introducing an additional phase $\varphi _{\mathit{X}\left ( \mathit{Y} \right )} =0$ or $\pi$ ($\frac{\pi }{2}$ or $\frac{3\pi }{2}$) in the $l$ pulses.
	Correspondingly, the states of $\mathit{X}$ basis can be written as $\left | + \right \rangle = \frac{\left ( \left | 0  \right \rangle + \left | 1  \right \rangle \right )}{\sqrt{2}}$ and $\left | - \right \rangle = \frac{\left ( \left | 0  \right \rangle - \left | 1  \right \rangle \right )}{\sqrt{2}}$, and the states in the $\mathit{Y}$ basis are $\left | +i \right \rangle = \frac{\left ( \left | 0  \right \rangle + i\left | 1  \right \rangle \right )}{\sqrt{2}}$ and $\left | -i \right \rangle = \frac{\left ( \left | 0  \right \rangle - i\left | 1  \right \rangle \right )}{\sqrt{2}}$. When the $\mathit{Z}$ basis is selected for encoding, the next Sagnac-IM modulates the $\left | 0  \right \rangle$ or $\left | 1  \right \rangle$ states by suppressing the $l$ or $s$ pulses. For the $\mathit{X}$ or $\mathit{Y}$ basis, both the $l$ and $s$ pulses pass through Sagnac-IM, which adjusted the average intensity of the $\mathit{X}$ or $\mathit{Y}$ basis to match that of the $\mathit{Z}$ basis.
	Subsequently, the light pulses are sent to Bob through a standard commercial single-mode fiber channel.
	
	After the light pulses reach the receiver (Bob), a polarization controller (PC) actively compensates for the polarization drift caused by the channel,  the compensation method is shown in Appendix~\ref{Appendix C1}.
	Subsequently, a PM is used to actively select the measurement basis, applying random phases of $0$ for the $\mathit{X}$ basis or $\frac{\pi }{2}$ for the $\mathit{Y}$ basis to the $l$ pulses.	The signals are then interfered in an asymmetric Mach-Zehnder interferometer (AMZI) with the same delay as the AMZI in the transmitter.
	
	The interfered states are detected by a superconducting nanowire single-photon detector (SNSPD; P-SPD8S, Photoec, China) with a detection efficiency of approximately 70\%, a dark count rate of about 20 Hz. The detection events are recorded using a time-to-digital converter (TDC) with  a gate width of 800 ps. Due to the limited availability of equipment, all devices, including the laser, PMs and TDC, are triggered and synchronized by using a high-speed arbitrary waveform generator with sampling rate of 2.4 GSa/s (HDAWG8, Zurich Instruments). \blk
	
	\section{Experimental results}
	\label{Experimental results}
	
	We perform a series of one-decoy BB84 RFI QKD over a commercial fiber spool to test the setup. For  each distance, we preset a reference-frame misalignment angle $\theta=\pi /9 $ (see Appendix~\ref{Appendix C} for a detailed method), and  a total of $N_{tot}=8.1\times 10^{11} $ pulses are sent out. Furthermore, we optimize implementation parameters, including the intensity, the probability of different intensities, and the probability of different bases  by establishing the system model~\cite{Li2020,Zhang2018} and applying a genetic algorithm to enhance the search speed~\cite{Li2022_GA} (see Appendix~\ref{Appendix C2} for a detailed procedure). \blk For instance, at 250 km, the signal and decoy state intensities sent after optimization are $\mu=0.388$ and $\nu=0.123$, respectively. The probabilities of the signal and decoy state are $P_{\mu}=0.5$ and $P_{\nu}=0.5$, respectively. Similarly, the probabilities of the $\mathit{Z}$, $\mathit{X}$, and $\mathit{Y}$ basis are $P_{Z_{A}}=0.476$, $P_{X_{A}}=0.262$, and $P_{Y_{A}}=0.262$, respectively. Bob performs basis selection with probabilities $P_{Z_{B}}=0.476$, $P_{X_{B}}=0.262$, and $P_{Y_{B}}=0.262$. 
	
	\begin{figure}[h]
		\centering
		\includegraphics[width=1\linewidth]{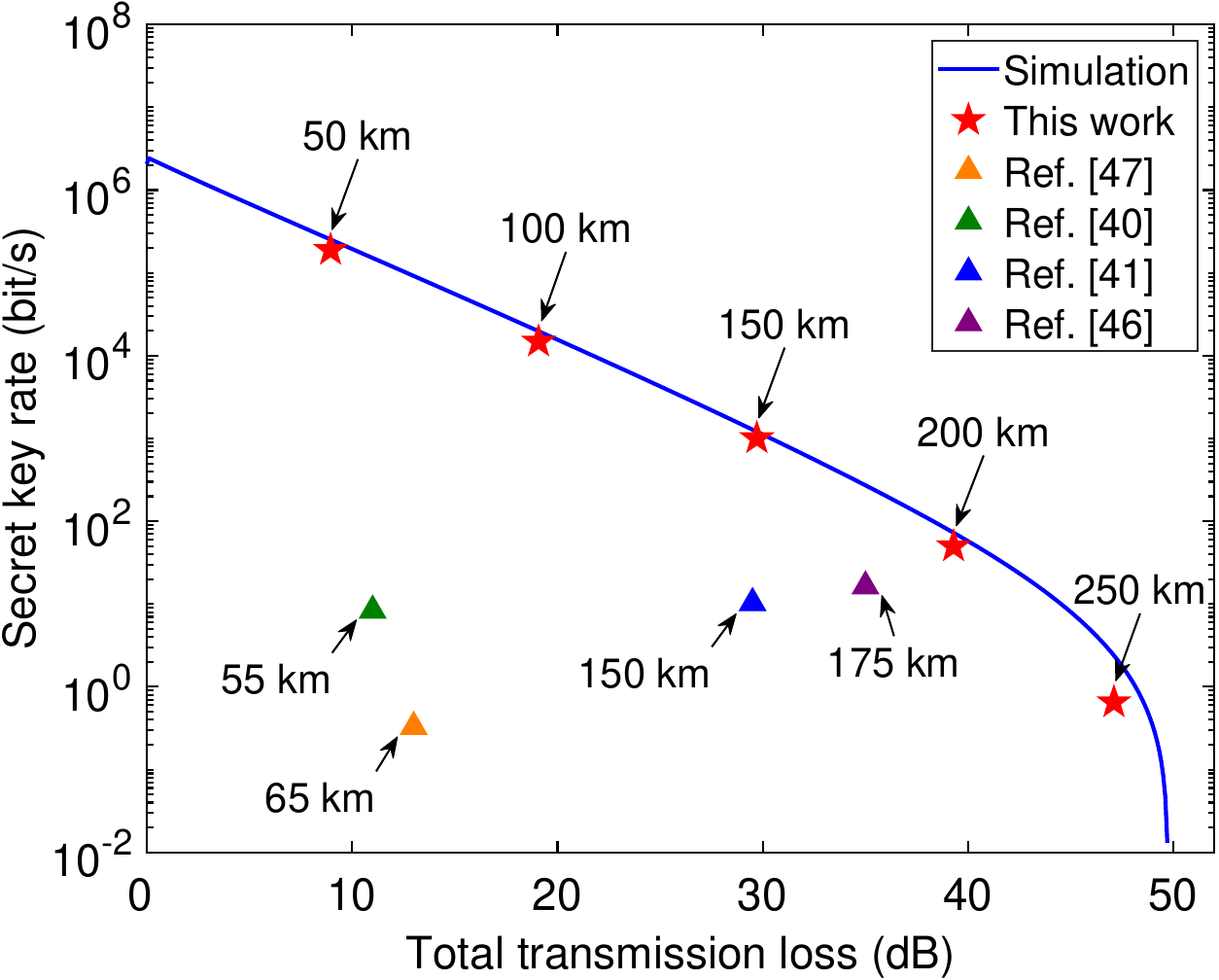}
		\caption{SKRs with different total transmission loss. The red five-pointed star symbolizes the experimental results, while the blue line represents the theoretical simulation results obtained based on the experimental parameters. The total transmission loss at 50, 100, 150 km, 200, and 250 km are measured as 8.95, 19.08, 29.71, 39.29, and 47.10 dB, respectively.   The triangles in different colors represent the results of state-of-the-art RFI QKD experiments.}
		\label{Key_Rate}
	\end{figure}	
	
	As shown in Fig.~\ref{Key_Rate}, the experimental demonstration is achieved at 50-, 100-, 150-, 200-, and 250-km optical fiber, with the reference frame misalignment angles, estimated from the experimental data, fluctuating around $\theta=\pi /9 $ during the tests. \blk The experimental results closely match the theoretical simulations, and the detailed experimental results are shown in Appendix~\ref{Appendix D}. We successfully perform key distributions over a distance of 250 km,  with the QBER of the system remaining around 1.77\%, generating a SKR of approximately 0.65 bit/s.
	
	To illustrate the progress entailed by our results, as shown in Tab.~\ref{Comparison}, we compare the experimental results of our system with those of state-of-the-art RFI QKD systems. It can been seen that our work represents the highest  reported key rate, longest distance.  Specifically, the SKR at 150 km is 100 times higher than the highest recorded SKR achieved reported in Ref.~\cite{Tian2024}.
	
	\section{Conclusions and discussion}
	\label{Conclusions and discussion}
	In conclusion, we have reported a RFI QKD system with a repetition rate of 150 MHz. The experimental results show that our system can maintain a low QBER and produce a respectable SKR when considering the misalignment of reference frame and finite-key effects. We achieved secure key bits distribution over a 250 km of optical fiber, which is the longest transmission distance of RFI QKD ever recorded. 
	
    As the next step, RFI QKD systems with GHz repetition rates and communication distances exceeding 400 km can be achieved by developing state-of-the-art electronic technology~\cite{Grunenfelder2023,Li2023} and using ultralow dark-count detectors ~\cite{Boaron2018,Liu2023}. \blk Moreover, to address the challenges posed by a more complex practical environment, we can explore the utilization of qubit-synchronization technology~\cite{Luca2020,Huang2024,Chen2024} to eliminate external electrical synchronization signals, and investigate the free-running RFI QKD scheme~\cite{Tang2022} to cope with the rapid variation of reference frames. 
	
	\section*{Acknowledgments}
	This study was supported by the National Natural Science Foundation of China (Nos. 62171144, 62031024, and  11865004), Guangxi Science Foundation (Nos.2021GXNSFAA220011 and 2021AC19384), and the Open Fund of IPOC (BUPT) (No. IPOC2021A02). 
	
	\appendix 
	\section{ONE-DECOY STATE RFI-QKD PROTOCOL}
	\label{Appendix A}
	In the one-decoy state RFI QKD protocol~\cite{Li2020}, Alice prepares quantum states of intensity $k\in \left \{ \mu ,\nu  \right \} $ on three orthogonal bases $\alpha \in \left \{ Z , X, Y \right \} $, which are then measured by Bob with bases $\beta \in \left \{ Z , X, Y \right \} $. Alice and Bob generate raw key bits using $\mathit{Z}$ basis, and estimate Eve’s information using $\mathit{X}$ and $\mathit{Y}$ bases.
	
	After the quantum channel transmission, the gain with intensity $k$ of quantum states that Alice sends in $\alpha$ basis and Bob measures in $\beta$ basis can be given by~\cite{Zhang2018}
	\begin{equation}
		Q_{\alpha \beta}^k=\frac{1}{2}\left(P_{\alpha^0 \beta^0}^k+P_{\alpha^0 \beta^1}^k+P_{\alpha^1 \beta^0}^k+P_{\alpha^1 \beta^1}^k\right), \label{Q}
	\end{equation}
	where $\alpha \beta\in \left \{ Z Z,~X X,~X Y,~Y X,~Y Y \right \} $, $\frac{1}{2}  $ represents the probability of Alice preparing quantum states $\alpha^0$ or $\alpha^1$. $\alpha^0$ $(\beta^0)$ and $\alpha^1$ $(\beta^1)$ consist of basis $\alpha$ $(\beta)$. $P_{\alpha^0 \beta^0}^k$ represents the probability that Bob measures $\beta^0$ when Alice prepares the quantum state $\alpha^0$ with intensity $k$, and the definition of  $P_{\alpha^0 \beta^1}^k$, $P_{\alpha^1 \beta^0}^k$, and $P_{\alpha^1 \beta^1}^k$ are similar. The  explicit expressions of $P_{\alpha^0 \beta^0}^k$, $P_{\alpha^0 \beta^1}^k$, $P_{\alpha^1 \beta^0}^k$, and $P_{\alpha^1 \beta^1}^k$ are as follows:
	\begin{equation}
		\begin{aligned}
			& P_{Z^0 Z^0 \left ( Z^1 Z^1 \right ) }^k=e^{-k \eta}(1-p_{d} )\left(e^{k \eta}+p_{d} -1\right), \\
			& P_{Z^0 Z^1 \left ( Z^1 Z^0 \right )}^k=e^{-k \eta}(1-p_{d}) p_{d} ,
		\end{aligned} \label{P_ZZ}
	\end{equation}
	
	\begin{equation}
		\begin{aligned}
			& P_{X^0 X^0 \left ( X^1 X^1\right ) }^k=e^{-k \eta}(1-p_{d} )\left(e^{k \eta \frac{1+\cos \theta}{2}}+p_{d} -1\right), \\
			& P_{X^0 X^1 \left ( X^1 X^0\right )}^k=e^{-k \eta}(1-p_{d})\left(e^{k \eta \frac{1-\cos \theta}{2}}+p_{d}-1\right),
		\end{aligned} \label{P_XX}
	\end{equation}
	
	\begin{equation}
		\begin{aligned}
			& P_{X^0 Y^0 \left ( X^1 Y^1\right ) }^k=e^{-k \eta}(1-p_{d})\left(e^{k \eta \frac{1-\sin \theta}{2}}+p_{d}-1\right), \\
			& P_{X^0 Y^1 \left ( X^1 Y^0\right )}^k=e^{-k \eta}(1-p_{d})\left(e^{k \eta \frac{1+\sin \theta}{2}}+p_{d}-1\right),
		\end{aligned} \label{P_XY}
	\end{equation}
	
	\begin{equation}
		\begin{aligned}
			& P_{Y^0 X^0 \left ( Y^1 X^1\right ) }^k=e^{-k \eta}(1-p_{d})\left(e^{k \eta \frac{1+\sin \theta}{2}}+p_{d}-1\right), \\
			& P_{Y^0 X^1 \left ( Y^1 X^0\right )}^k=e^{-k \eta}(1-p_{d})\left(e^{k \eta \frac{1-\sin \theta}{2}}+p_{d}-1\right),
		\end{aligned} \label{P_YX}
	\end{equation}
	
	\begin{equation}
		\begin{aligned}
			& P_{Y^0 Y^0 \left ( Y^1 Y^1\right ) }^k=e^{-k \eta}(1-p_{d})\left(e^{k \eta \frac{1+\cos \theta}{2}}+p_{d}-1\right), \\
			& P_{Y^0 Y^1 \left ( Y^1 Y^0\right )}^k=e^{-k \eta}(1-p_{d})\left(e^{k \eta \frac{1-\cos \theta}{2}}+p_{d}-1\right),
		\end{aligned} \label{P_YY}
	\end{equation}
	where $p_{d}$ is dark-count rate. $\eta=\eta_{d}\times 10^{-\frac{loss}{10}}$ is overall transmission efficiency. $\eta_{d}$ is detection efficiency. "loss" is the total transmission loss. $\theta$ is the misalignment angle of reference frames between Alice and Bob.
	
	With Eqs.~(\ref{Q})-(\ref{P_YY}), the expressions for $Q_{\alpha \beta}^k$ can be obtained as
	\begin{equation}
		\begin{aligned}
			Q_{Z Z}^k=e^{-k \eta}(1-p_{d})\left(e^{k \eta}+2 p_{d}-1\right),
		\end{aligned} \label{Q_ZZ}
	\end{equation}
	
	\begin{equation}
		\begin{aligned}
			Q_{X X}^k=e^{-k \eta}(1-p_{d})\left(e^{k \eta \frac{1+\cos \theta}{2}}+e^{k \eta \frac{1-\cos \theta}{2}}+2 p_{d}-2\right),
		\end{aligned} \label{Q_XX}
	\end{equation}
	
	\begin{equation}
		\begin{aligned}
			Q_{X Y}^k=e^{-k \eta}(1-p_{d})\left(e^{k \eta \frac{1+\sin \theta}{2}}+e^{k \eta \frac{1-\sin \theta}{2}}+2 p_{d}-2\right),
		\end{aligned} \label{Q_XY}
	\end{equation}
	
	\begin{equation}
		\begin{aligned}
			Q_{Y X}^k=e^{-k \eta}(1-p_{d})\left(e^{k \eta \frac{1+\sin \theta}{2}}+e^{k \eta \frac{1-\sin \theta}{2}}+2 p_{d}-2\right),
		\end{aligned} \label{Q_YX}
	\end{equation}
	
	\begin{equation}
		\begin{aligned}
			Q_{Y Y}^k=e^{-k \eta}(1-p_{d})\left(e^{k \eta \frac{1+\cos \theta}{2}}+e^{k \eta \frac{1-\cos \theta}{2}}+2 p_{d}-2\right).
		\end{aligned} \label{Q_YY}
	\end{equation}
	
	Considering the optical intrinsic error rate $e_{d}$, the QBER with intensity $k$ in $\alpha \beta$ basis $(E_{\alpha \beta}^k)$ can be denoted as 
	\begin{equation}
		\begin{aligned}
			E_{\alpha \beta}^k=\min \left\{\tilde{E}_{\alpha \beta}^k, 1-\tilde{E}_{\alpha \beta}^k\right\},
		\end{aligned} \label{E}
	\end{equation}
	where $\tilde{E}_{\alpha \beta}^k=e_d\left(1-2 e_{\alpha \beta}^k\right)+e_{\alpha \beta}^k$, and $e_{\alpha \beta }^{k}=\frac{P_{\alpha ^{0} \beta ^{1}}^{k}+P_{\alpha ^{1} \beta ^{0}}^{k}}{2 Q_{\alpha \beta }^{k}}$. Without loss of generality, we assume $E_{\alpha \beta}^k\le 0.5$ in Eq.~(\ref{E}). Otherwise, Alice or Bob flips her or his bit strings to make $E_{\alpha \beta}^k\le 0.5$.
	
	Through the above formulas, the number of detections $(n_{\alpha \beta}^k)$ and error detections $(m_{\alpha \beta}^k)$ with the intensity $k$ in $\alpha \beta$ basis can be calculated
	\begin{equation}
		\begin{aligned}
			n_{\alpha \beta}^{k} = N_{tot}P_{\alpha \beta }^{k\text{,det}} Q_{\alpha \beta}^{k},
		\end{aligned} \label{n}
	\end{equation}
	
	\begin{equation}
		\begin{aligned}
			m_{\alpha \beta}^{k} = N_{tot}P_{\alpha \beta }^{k\text{,det}} Q_{\alpha \beta}^{k}E_{\alpha \beta}^k,
		\end{aligned} \label{m}
	\end{equation}
	where $N_{tot}$ is the total pulse numbers sent by Alice. $P_{\alpha \beta }^{k\text{,det}} = P_{\alpha} P_{\beta} P_{k}$, $P_{\alpha}$ is the probability that Alice sends $\alpha$ basis, $P_{\beta}$ is the  probability that Bob chooses $\beta$ basis measurement, and $P_{k}$ is the probability that Alice sends intensity $k$.
	
	Considering the finite-key effects~\cite{Rusca2018,Lim2014}, according to the Hoeffding inequality, we can get the upper and lower bounds of $n_{\alpha \beta}^k$ $(m_{\alpha \beta}^k)$ as follows:
	\begin{equation}
		\begin{array}{l}
			n_{\alpha  \beta }^{k, \pm}:=\frac{e^{k}}{P_{k}}\left(n_{\alpha \beta }^{k} \pm \sqrt{\frac{n_{\alpha \beta }}{2} \log  \frac{1}{\varepsilon_{1}}}\right) , \\
		\end{array} \label{n_UL}
	\end{equation}
	
	\begin{equation}
		\begin{array}{l}
			m_{\alpha  \beta }^{k, \pm}:=\frac{e^{k}}{P_{k}}\left(m_{\alpha \beta }^{k} \pm \sqrt{\frac{m_{\alpha \beta }}{2} \log  \frac{1}{\varepsilon_{2}}}\right) . \\
		\end{array} \label{m_UL}
	\end{equation}
	
	The lower bound $(L)$ of single-photon events and vacuum events, and the upper bound $(U)$ of vacuum events in $\alpha  \beta$ basis can be estimated as
	
	\begin{equation}
		\begin{array}{l}
			s_{\alpha \beta, 1}^{L}=\frac{\tau_{1} \mu}{\nu (\mu-\nu )}\left(n_{\alpha \beta }^{\nu ,-}-\frac{\nu ^{2}}{\mu^{2}} n_{\alpha \beta}^{\mu,+}-\frac{\mu^{2}-\nu ^{2}}{\mu^{2}} \frac{s_{\alpha \beta, 0}^{U}}{\tau_{0}}\right), \\
		\end{array} \label{s_1L}
	\end{equation}
	
	\begin{equation}
		\begin{array}{l}
			s_{\alpha \beta ,0}^{L} =\frac{\tau _{0} }{\mu -\nu } \left ( \mu n_{\alpha \beta }^{\nu,-} - \nu  n_{\alpha \beta }^{\mu,+} \right ), \\
		\end{array} \label{s_0L}
	\end{equation}
	
	\begin{equation}
		\begin{aligned}
			s_{\alpha \beta ,0}^{U} &=\text{min} \left \{ 2\left [ m_{\alpha \beta } +\delta \left ( n_{\alpha \beta },\varepsilon _{1}   \right ) \right ] , \right. \\
			& \left. 2\tau_{0} \frac{e^{k} }{P_{k} } \left [ m_{\alpha \beta }^{k} + \delta \left ( m_{\alpha \beta },\varepsilon _{2} \right ) \right ]+2\delta \left ( n_{\alpha \beta },\varepsilon _{1} \right ) \right \}, \\
		\end{aligned} \label{s_0U}
	\end{equation}
	where $\tau _{i} =\sum_{k\in \left \{ \mu ,\nu  \right \} }  P_{k}e^{-k} \frac{k^{i} }{i!} $ is the total probability to send an $i$-photon state.
	
	The upper bound of single-photon error rate $e_{\alpha \beta}^{1,U}$, and the upper bound of the single-photon bit errors $m_{\alpha \beta,bit}^{1,U}$ in $\alpha  \beta$ basis can be denoted as
	\begin{equation}
		\begin{array}{l}
			e_{\alpha \beta}^{1,U} =\text{min}\left \{\frac{m_{\alpha \beta,bit}^{1,U}}{s_{\alpha \beta,1}^{L}},~0.5 \right \}, \\
		\end{array} \label{e_1U}
	\end{equation}
	
	\begin{equation}
		\begin{array}{l}
			m_{\alpha \beta, b i t}^{1, U}= \frac{\tau_{1}}{\mu-\nu}\left(m_{\alpha \beta}^{\mu,+}-m_{\alpha \beta}^{\nu,-}\right). \\
		\end{array} \label{e_1Ubit}
	\end{equation}

    \section{EXPERIMENTAL DETAILS}
	\label{Appendix B}
	\subsection{Polarization compensation}  \label{Appendix C1}
	For our RFI QKD system, although no phase-feedback devices are required, polarization drift decreases the system's secret key rate. Specifically, for the $Z$ basis, polarization drift decreases the number of detected counts due to the polarization dependence of the superconducting nanowire single-photon detector, leading to a lower secret key rate. For the $X$ and $Y$ bases, polarization drift may cause different polarizations in the early and late pulses, reducing the visibility of interference fringes, increasing the quantum bit-error rate, and ultimately lowering the system's secret key rate.

	To eliminate the influence of polarization drift,  we actively compensate for the polarization drift caused by the channel using a mechanical three-ring polarization controller (PC). The specific compensation method can be summarized in the following three steps:
	\newline
	Step 1: The $\mathit{Z}$ basis count ($n_{ZZ}$) is measured, and $n_{ZZ}$ is maximized by adjusting the PC.
	\newline
	Step 2: The error rate of $\mathit{X}$ basis and $\mathit{Y}$ basis ($E_{XX}$ and $E_{YY}$) are estimated from the experimental results, and $E_{XX}$ and $E_{YY}$ are minimized by fine tuning the PC.
	\newline
	Step 3: Repeat the above two steps to maximize $n_{ZZ}$ while minimizing $E_{XX}$ and $E_{YY}$.

	\subsection{Parameter optimization}  \label{Appendix C2}
	For better system performance, it is advantageous to optimize operating parameters $\vec{\upsilon}$, e.g., signal- and decoy-state intensities ($\mu$ and  $\nu$), the probabilities of the intensities ($P_{\mu}$ and $P_{\nu}$), and the probabilities of the bases ($P_{Z_{A}}=P_{Z_{B}}$, $P_{X_{A}}=P_{X_{B}}$, and $P_{Y_{A}}=P_{Y_{B}}$). 
	
	To calculate the secret key rate of our RFI QKD system, we must consider the influence of experimental parameters, including detection efficiency $\eta_{d}$, the dark-count rate $p_{d}$, the optical intrinsic error rate $e_{d}$, error-correction efficiency $f$, distance $L$ (unit is dB), and total number of light pulses $N_{tot}$ sent by Alice. These experimental parameters are denoted as $\vec{e}$, so the secret-key-rate function of RFI QKD can be expressed as
	\begin{equation}
		\mathcal{F}=R(\vec{e}, \vec{\upsilon}),
	\end{equation}
	which is a function of operating parameters $\vec{\upsilon}$ and experimental parameters $\vec{e}$. The experimental parameters are typically fixed, as determined by the performance of RFI QKD systems,	and the operating parameters are controlled by the users. Therefore, to obtain the optimal secret key rate, we must calculate
	\begin{equation}
		R_{\max }(\vec{e})=\max _{\vec{\upsilon} \in S} R(\vec{e}, \vec{\upsilon}),
	\end{equation}
	where $S$ is the search space for the parameters. 
	
	The key objective of RFI QKD parameter optimization is to search the optimal set of $\vec{\upsilon}_{\text{opt}}$ that maximizes the objective function $R(\vec{e}, \vec{\upsilon})$ with the given $\vec{e}$, i.e., it can be viewed as searching for
	\begin{equation}
		\vec{\upsilon}_{\text {opt }}(\vec{e})=\underset{\vec{\upsilon} \in S}{\arg \max } R(\vec{e}, \vec{\upsilon}).
	\end{equation}
	
	Taking 250 km as an example, in our implementation, we predetermined the experimental parameters of the system, i.e., $\eta_{d}=70\%$, $p_{d}=1\times 10^{-8}$, $e_{d}^{Z} =0.7\%$, $e_{d}^{X(Y)} =1.4\%$, $f=1.16$,  $N_{tot}=8.1\times 10^{11} $ and $L= 47.10 $ dB. Then, these experimental parameters are substituted into the secret-key-rate formula of RFI-QKD (Eq.~(\ref{key rate})), and a genetic algorithm~\cite{Li2022_GA} is applied for optimization to obtain the optimal operating parameters according to the aforementioned method, i.e., $\mu=0.388$, $\nu=0.123$, $P_{\mu}=0.5$, $P_{\nu}=0.5$, $P_{Z_{A}}=P_{Z_{B}}=0.476$, $P_{X_{A}}=P_{X_{B}}=0.262$, and $P_{Y_{A}}=P_{Y_{B}}=0.262$.
	\blk
	
	\begin{figure*}[htbp]
		\subfigure{
			\label{EXXu}
			\includegraphics[height=180pt,width=230pt]{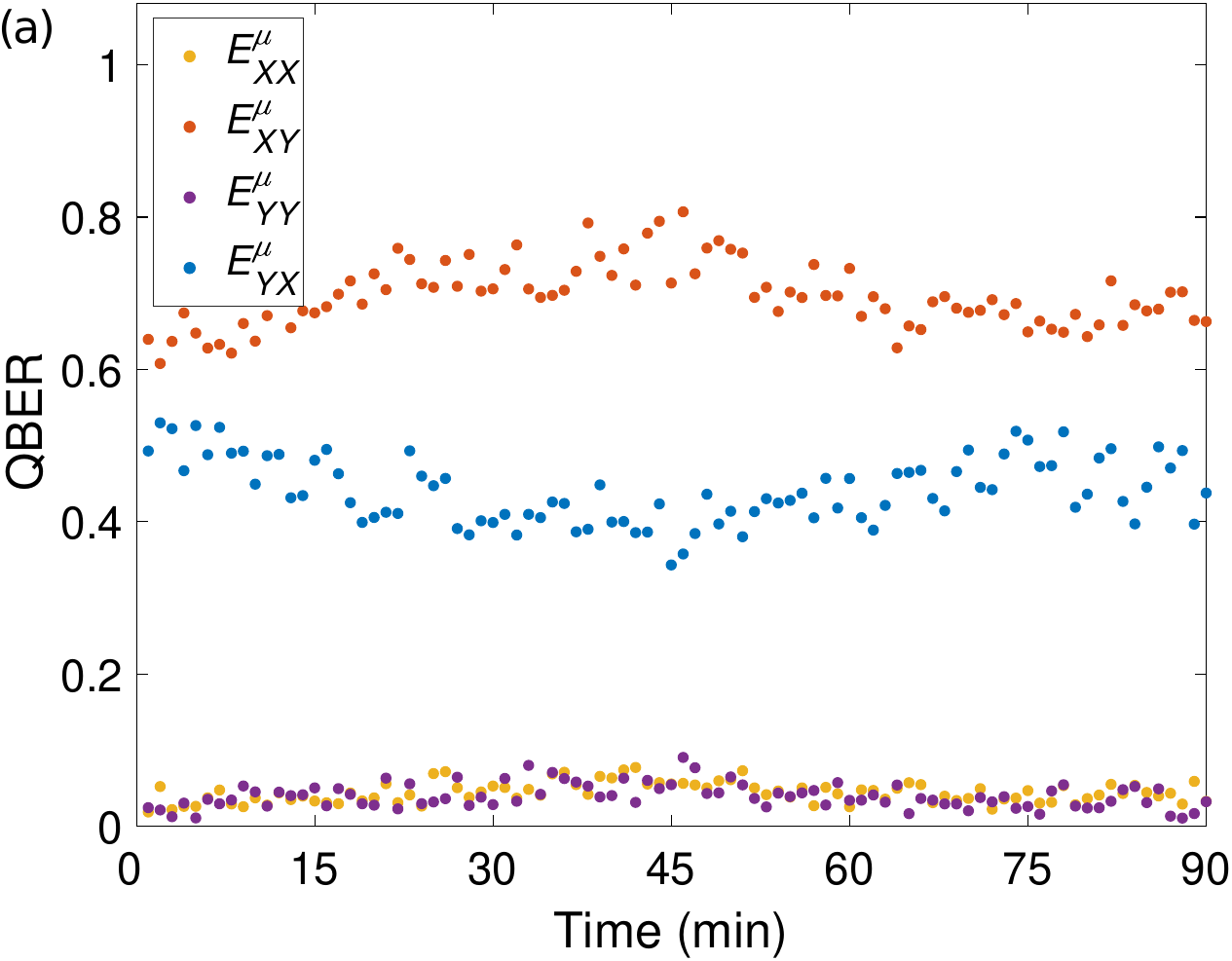}}
		\quad 
		\subfigure{
			\label{EXXv}
			\includegraphics[height=180pt,width=230pt]{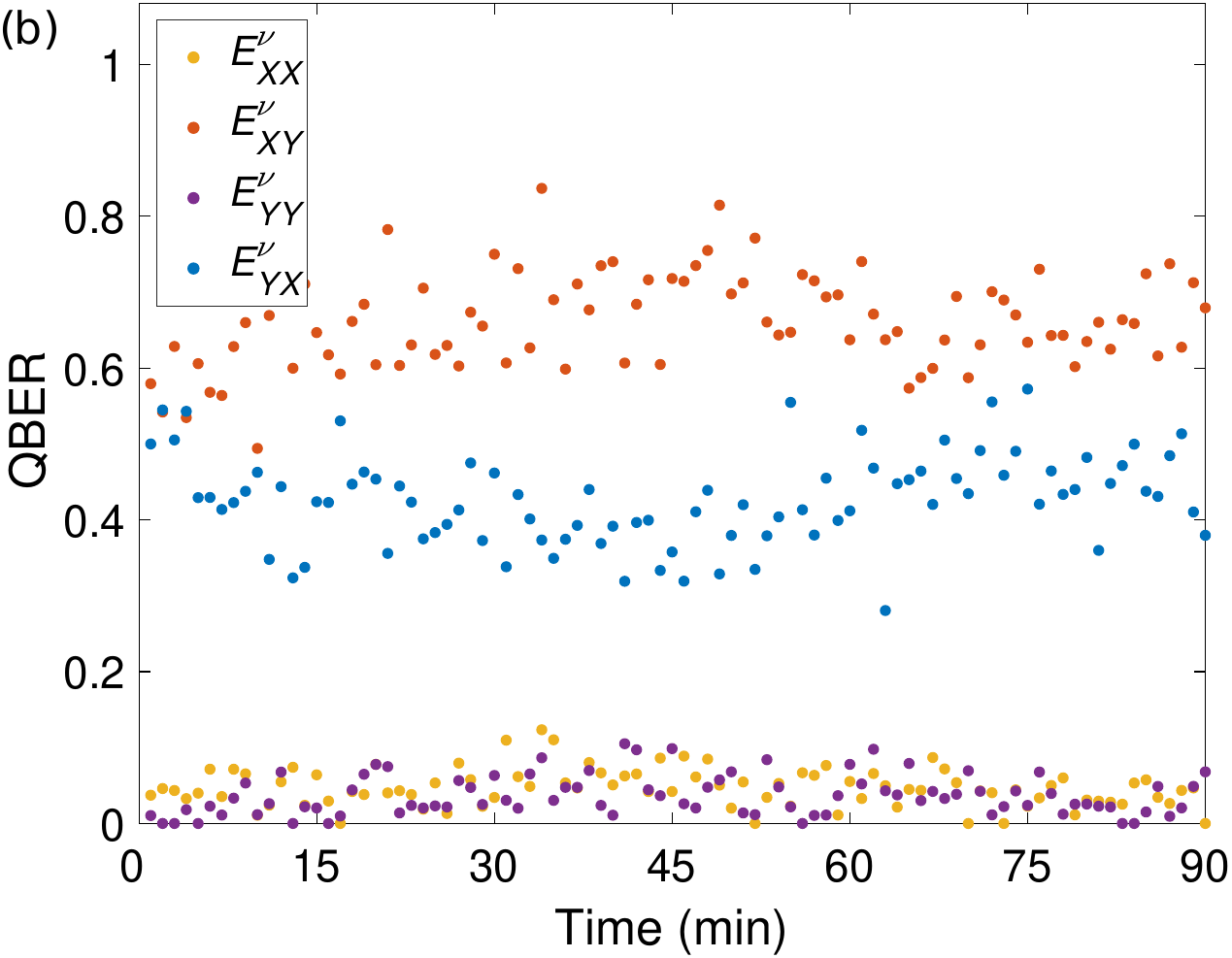}}
		
		\caption{QBERs measured on $\mathit{X}$ and $\mathit{Y}$ bases at 250 km. Each point is measured within 1 min. (a) The QBERs of signal states on $\mathit{X}$ and $\mathit{Y}$ bases. (b) The QBERs of decoy states on $\mathit{X}$ and $\mathit{Y}$ bases.}
		\label{QBERXY}
	\end{figure*}
	
	\begin{figure}[htbp]
		\centering
		\includegraphics[width=1\linewidth]{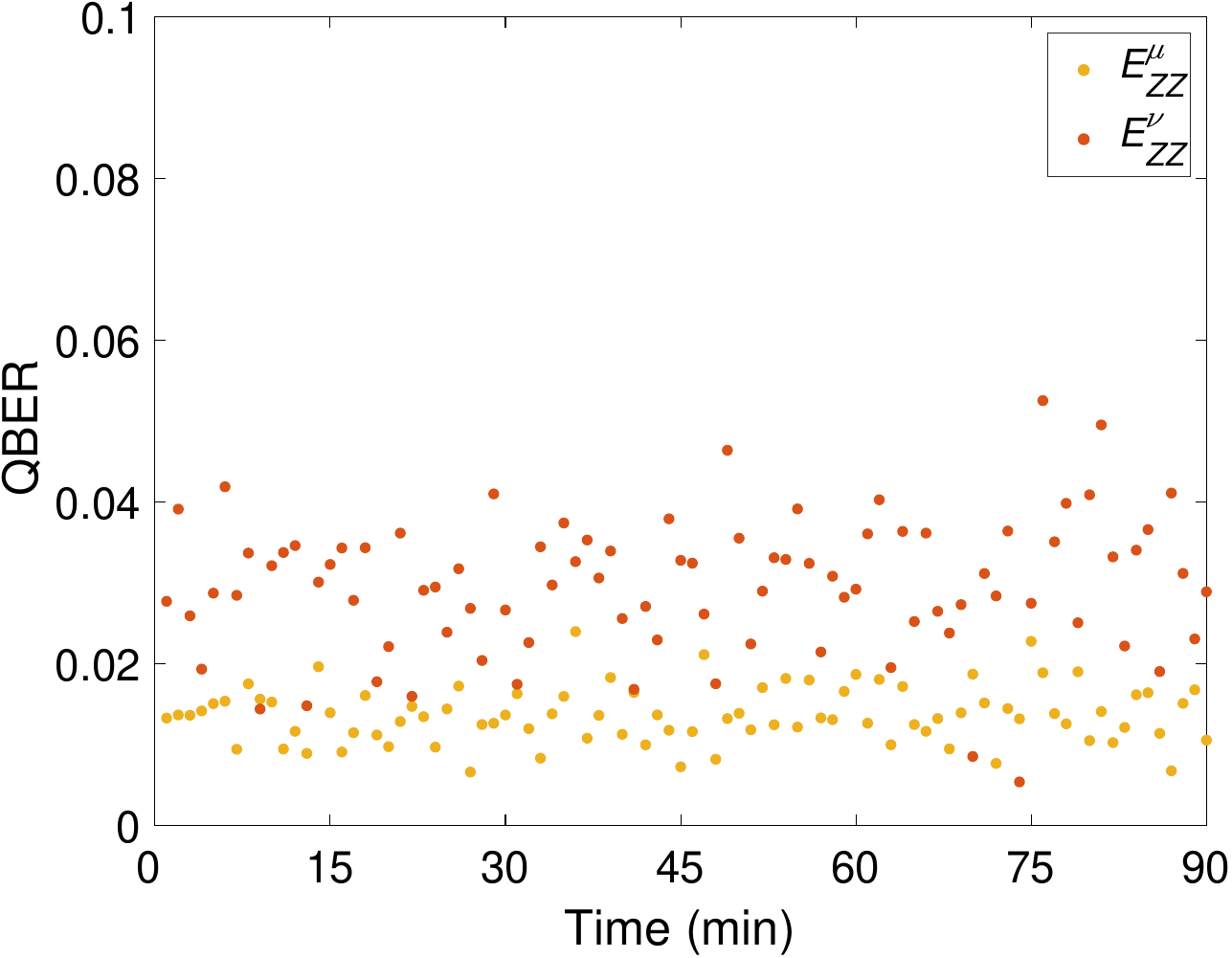}
		\caption{The QBERs of signal and decoy states on $\mathit{Z}$ basis at 250 km. Each point is measured within 1 min.}
		\label{QBERZZ}
	\end{figure}
	
	\begin{table*}[]
		\centering
		\caption{The experimental results for the different fiber lengths.}
		\fontsize{5}{9} \selectfont
		\setlength\tabcolsep{8pt}
		\resizebox{\linewidth}{!}{
			\begin{tabular}{cccccc}
				\hline \hline
				Fiber length (km) & 50            & 100          & 150         & 200        & 250 \\
				Total loss~(dB)   & 8.95          & 19.08        & 29.71       & 39.29      &47.10 \\ \hline
				$\mu$             & 0.468         & 0.441        & 0.417       & 0.399      & 0.388 \\
				$\nu$             & 0.073         & 0.091        & 0.108       & 0.115      & 0.123 \\
				$P_{\mu}$         & 0.847         & 0.769        & 0.647       & 0.510      & 0.500 \\
				$P_{\nu}$         & 0.153         & 0.231        & 0.353       & 0.490      & 0.500 \\
				$P_{Z_{A}}$       & 0.928         & 0.894        & 0.834       & 0.710      & 0.476 \\
				$P_{X_{A}}$       & 0.036         & 0.053        & 0.083       & 0.145      & 0.262 \\
				$P_{Y_{A}}$       & 0.036         & 0.053        & 0.083       & 0.145      & 0.262 \\
				$P_{Z_{B}}$       & 0.928         & 0.894        & 0.834       & 0.710      & 0.476 \\
				$P_{X_{B}}$       & 0.036         & 0.053        & 0.083       & 0.145      & 0.262 \\
				$P_{Y_{B}}$       & 0.036         & 0.053        & 0.083       & 0.145      & 0.262 \\
				$n_{Z Z}^{\mu}$   & 3813901712    & 305661861    & 21117282    & 1115911    & 93130 \\
				$n_{Z Z}^{\nu}$   & 108348661     & 19011000     & 2960939     & 300311     & 31187 \\
				$n_{X X}^{\mu}$   & 5184273       & 1119185      & 219192      & 40320      & 27506 \\
				$n_{X X}^{\nu}$   & 181984        & 75763        & 30141       & 13485      & 8502 \\
				$n_{X Y}^{\mu}$   & 6282138       & 915526       & 210517      & 48374      & 27116 \\
				$n_{X Y}^{\nu}$   & 169633        & 68722        & 30062       & 12152      & 8635 \\
				$n_{Y X}^{\mu}$   & 5818916       & 918488       & 212863      & 37934      & 26864 \\
				$n_{Y X}^{\nu}$   & 168317        & 75229        & 26784       & 13779      & 8251 \\
				$n_{Y Y}^{\mu}$   & 6165481       & 1171963      & 207771      & 45700      & 27782 \\
				$n_{Y Y}^{\nu}$   & 165149        & 76967        & 31728       & 13177      & 8334 \\ \hline
				$m_{Z Z}^{\mu}$   & 35279201      & 2464002      & 158010      & 11638      & 1274 \\
				$m_{Z Z}^{\nu}$   & 1187951       & 188281       & 29600       & 5680       & 924 \\
				$m_{X X}^{\mu}$   & 222297        & 52369        & 9390        & 1758       & 1224 \\
				$m_{X X}^{\nu}$   & 9400          & 3796         & 1478        & 501        & 403 \\
				$m_{X Y}^{\mu}$   & 2473288       & 306729       & 69422       & 19491      & 8241 \\
				$m_{X Y}^{\nu}$   & 74943         & 24423        & 11025       & 6016       & 2930 \\
				$m_{Y X}^{\mu}$   & 1703022       & 361376       & 84665       & 13279      & 11855 \\
				$m_{Y X}^{\nu}$   & 51575         & 30302        & 10149       & 4652       & 3529 \\
				$m_{Y Y}^{\mu}$   & 266791        & 58116        & 9526        & 1948       & 1095 \\
				$m_{Y Y}^{\nu}$   & 9461          & 3980         & 1319        & 759        & 307 \\ \hline
				$C$               & 1.34          & 1.24         & 1.16        & 1.20       & 0.78 \\
				$e_{Z Z}^{1,U}$   & 1.60\%        & 1.38\%       & 1.30\%      & 1.82\%     & 2.47\% \\
				$E_{Z Z}$         & 0.93\%        & 0.82\%       & 0.78\%      & 1.22\%     & 1.77\% \\
				$s_{Z Z, 0}^{L}$  & $10^{-10}$    & $10^{-10}$   & $10^{-10}$  & $10^{-10}$ & $10^{-10}$ \\
				$s_{Z Z, 1}^{L}$  & 2380929352.24 & 202768408.33 & 15101596.85 & 844518.85  & 71757.71 \\
				$SKR$ (bit/s)         & 189080.80     & 14773.20     & 1004.15     & 49.65      & 0.65 \\ \hline \hline
			\end{tabular} 
			\label{Table1}
		}
	\end{table*}
	
	\section{THE DERIVATION OF REFERENCE FRAME MISALIGNMENT}
	\label{Appendix C}
	In the practical RFI QKD scheme, the QBER of the $\mathit{X}$ ($\mathit{Y}$) basis is affected by the dark count rate $p_{d}$, the optical intrinsic error rate $e_{d}$, and the reference-frame misalignment $\theta$. According to the Supplementary Material in~\cite{Tang2022}, when $\alpha \beta \in \left \{ XX,YY \right \} $, the QBER $e_{\alpha \beta }^{\mu}$ is calculated as 
	\begin{equation}
		e_{\alpha \beta}^{\mu}=\frac{e^{\mu \eta(1-\cos \theta) / 2}+p_{d}-1}{e^{\mu \eta(1-\cos \theta) / 2}+e^{\mu \eta(1+\cos \theta) / 2}+2 p_{d}-2}.
		\label{e_angle}
	\end{equation}
	When $\mu \eta(1-\cos \theta) / 2\ll 1$ and $\mu \eta(1+\cos \theta) / 2\ll 1$, $e_{\alpha \beta}^{\mu}$ can be simplified as
	\begin{equation}
		\begin{aligned}
			e_{\alpha \beta}^{\mu} & \approx \frac{\mu \eta(1-\cos \theta) / 2+p_{d}}{\mu \eta(1-\cos \theta) / 2+\mu \eta(1+\cos \theta) / 2+2p_{d}} \\
			& =1 / 2-\cos \theta /\left(2+4 p_{d} / \mu \eta\right),
		\end{aligned}
	\end{equation}
	where $p_{d}\ll \mu \eta$, $e_{\alpha \beta}^{\mu}$ can be further simplified as
	\begin{equation}
		e_{XX}^{\mu}\approx \frac{\left ( 1-cos\theta  \right ) }{2}. 
	\end{equation}
	
	Therefore, the reference-frame misalignment angle $\theta$ can be estimated from the experimentally measured QBER $E_{XX}^{\mu}$ as 
	\begin{equation}
		\theta =\arccos \left ( 1-2E_{XX}^{\mu} \right ) . \label{angle}
	\end{equation}

    Using above model, we can preset the reference-frame misalignment angle to approximately \(\pi/9\) by preprocessing accumulated experimental data from a 5-min period. Then let the system operate freely. \blk
    
	\section{DETAILED EXPERIMENTAL PARAMETERS AND RESULTS}
	\label{Appendix D}
	
	Table~\ref{Table1} summarizes the parameters employed in our experiment, which are the optimized values. Fiber length represents the lengths of the fibers between Alice and Bob in the experimental test, and total loss is the total loss of the corresponding fiber length. $\mu$ and $\nu$ are the intensity of signal state and decoy state, respectively. Due to the large jitter of the detector's falling edge at high count rates, the system's QBER is significantly affected. To mitigate this effect, we reduced the intensities by half for each distance. Furthermore, $P_{\mu}$ and $P_{\nu}$ are the probability that Alice sends intensity $\mu$ and $\nu$, respectively. Alice sends $\mathit{Z}$ $(\mathit{X},~\mathit{Y})$ basis at the probability of $P_{Z_{A}}$ $(P_{X_{A}},~P_{Y_{A}})$, while Bob actively selects the probability of $\mathit{Z}$ $(\mathit{X},~\mathit{Y})$ basis measurement to be $P_{Z_{B}}$ $(P_{X_{B}},~P_{Y_{B}})$.
	
	We strictly followed the steps of the decoy-state BB84 QKD protocol, where the intensities and encoded states of quantum signals are randomly modulated and measured. All counts were obtained using a standard postprocessing program to analyze the detection results~\cite{Grunenfelder2023,Li2023}. \blk We characterized our experimental system and then tested it experimentally with various fiber spools. The experimental results after measurement and postprocessing are summarized in Tab.~\ref{Table1}. $n_{\alpha \beta}^{k}$ and $m_{\alpha \beta}^{k}$ represents the number of detections and error detections with the intensity $k$ in $\alpha \beta$ basis, respectively. $C$ is an intermediate quantity used to estimate Eve's information, $e_{Z Z}^{1,U}$ is the upper bound of the single-photon error rate with $\mathit{Z Z}$ basis, and $E_{Z Z}$ is the QBER with $\mathit{Z Z}$ basis. $s_{Z Z, 0}^{L}$ and $s_{Z Z, 1}^{L}$ are the lower bound for counts of vacuum events and single-photon events with $\mathit{Z Z}$ basis estimated by the decoy-state method based on experimental results. SKR is the secret key rate obtained by the experiment. 
	
	Finally, we present the QBERs of the signal and decoy state for each basis at 250-km optical fiber in Fig.~\ref{QBERXY} and Fig.~\ref{QBERZZ}. During the system-testing process, no additional devices are required to provide phase feedback. Due to the relatively stable laboratory environment, the influence of environmental factors on the phase of the two interferometers is relatively small during the sampling period of one and a half hours, the reference frames experiences slow changes, leading to corresponding changes in the QBERs of $\mathit{X}$ and $\mathit{Y}$ bases (see Fig.~\ref{QBERXY}). According to the QBERs, the reference-frame misalignment angles can be estimated using Eq.~(\ref{angle}) to fluctuate around $\theta=\pi /9 $. \blk In contrast, $\mathit{Z}$ basis utilize time-bin phase coding, resulting in its QBER remaining relatively stable with little fluctuation (see Fig.~\ref{QBERZZ}). The QBER fluctuation of decoy state is larger than that of the signal state due to its susceptibility to statistical fluctuations. It is worthwhile to note that to mitigate the influence of dark counting, we set the detector gate width to 600 ps when collecting data at 250 km, with the detection efficiency of 70\%.

	\end{document}